\documentclass[english,prl,twocolumn]{revtex4}
\usepackage[T1]{fontenc}
\usepackage[latin1]{inputenc}
\usepackage{graphicx}
\usepackage{amssymb}

\makeatletter

\newcommand{\be}{\begin{equation}}
\newcommand{\ee}{\end{equation}}
\newcommand{\ve}{\epsilon}
\usepackage{babel}

\usepackage{babel}
\makeatother
\begin{document}

\title{Single-atom density of states of an optical lattice}

\begin{abstract}
We consider a single atom in an optical lattice, subject to a harmonic
trapping potential.  The problem is treated in the tight-binding approximation, with
an extra parameter $\kappa$ denoting the strength of the harmonic trap.
It is shown that the $\kappa \to 0$ limit of this problem is singular, in the
sense that the density of states for a very shallow trap
($\kappa \to 0$) is {\it qualitatively different\/} from that of 
a translationally invariant lattice ($\kappa = 0$).
The physics of this difference is discussed, and densities of states and
wave functions are exhibited and explained.
\end{abstract}

\author{Chris Hooley}

\affiliation{School of Physics and Astronomy, University of Birmingham,  
Edgbaston, Birmingham B15 2TT, U.K.}

\author{Jorge Quintanilla}

\affiliation{School of Physics and Astronomy, University of Birmingham,  
Edgbaston, Birmingham B15 2TT, U.K.}

\maketitle

The achievement of degeneracy in atomic gases has led to a surge of
activity. One reason is that these systems provide a test-bed for
theories developed in the context of solid-state physics.
Indeed there have been experimental observations of Bose-Einstein
condensation (BEC) \cite{Anderson:1995}, vortex lattices
\cite{AboShaeer:2001,Haljan:2001}, superfluid-insulator transitions
\cite{Greiner:2002}, fermion degeneracy \cite{DeMarco-Jin-99} and, very
recently, advances have been made towards Bardeen-Cooper-Schrieffer (BCS)
superfluidity \cite{Greiner-et-al-03}. A great
advantage of these systems is one's unprecedented control of the
experimental situation. By varying the magnetic field strength,
particle-particle interactions can be altered. Moreover, one can use
lasers to create an optical lattice whose parameters can be changed at will
\cite{Greiner-et-al-01}. The latter scenario has been the subject of
recent theoretical attention
\cite{Scott-et-al-02,Buchler-Blatter-Zwerger-02,Ho-Cazalilla-Giamarchi-03,Scott-Martin-et-al-03}.

One key difference between these systems and the usual models of the
solid state is that they are not translationally invariant, due to
the trapping potential necessary to confine the gas. However, in the
usual set-up, when the trap is the only external potential, the local
density approximation can be employed, and thermodynamic quantities
can be simply related to their values in the absence of the trap
\cite{Damle-Senthil-Majumdar-Sachdev-96}. Trapped atoms are,
therefore, a good model of the continuum many-particle system
--- analogous, in the fermion case, to `jellium', the ideal metal
of many-electron theory. One thus expects that, when
the periodic optical lattice potential is added, an experimental model of
electrons in real crystals can be obtained.

In this letter we take a first step towards testing this expectation by
obtaining the single-particle density of states (DOS). 
This is the starting point for understanding many key features of degenerate quantum systems. For example, in an ideal Bose gas it determines whether there is BEC \cite{Nozieres-95}, while in a weakly attractive Fermi gas it gives the critical temperature for BCS superfluidity \cite{DeGennes-66}. The DOS becomes a smooth curve only in the limit in which the trap is very weak, so that
the spectrum becomes a continuum. 
As we shall see, this limit
is singular, in the sense that the DOS is qualitatively different from
that of an infinite, translationally invariant lattice. 

\emph{Model.} We consider a single particle, with no internal degrees
of freedom, moving in an infinite tight-binding lattice with
one orbital per site and hopping between nearest neighbours. The
particle is confined to a finite region by a harmonic
trapping potential. For simplicity, we assume a $d$-dimensional
hypercubic lattice. For $d>1$, the problem is separable, so we shall
focus our attention on the 1D case, indicating how the results for $d
\geqslant 2$ follow as the need arises. The Hamiltonian for $d=1$
is
\begin{equation} H=-t\sum _{j}\left(\left|j\right\rangle \left\langle
j+1\right|+\textrm{H.c.}\right)+\frac{1}{2}\kappa \sum_{j}
\left(aj\right)^{2}\left|j\right\rangle \left\langle  
j\right|,\label{ham}
\end{equation}
where $j$ is a site label, $a$ the lattice constant, $\kappa $
the strength of the trap and $t$ the nearest-neighbour hopping
integral.
(This model also occurs in the related subject of the dynamical diffraction
of atoms by static light fields; see, for example, ref.~\cite{ODell-01} and
references therein.)
Note that the trap is centred at a lattice site; see
refs.~\cite{Chalbaud-Gallinar-Mata-86,Gallinar-Chalbaud-91} for a
discussion of the non-trivial effect of incommensuration. The model
can be generalised also by considering anisotropic lattices and
trapping potentials.

Before describing the properties of the model, let us discuss its validity.
In experiments \cite{Greiner-et-al-01}, the lattice is generated by
counter-propagating laser fields, giving rise to a static, periodic
potential \cite{Dahan-et-al-96} of the form $U(x)=-U_{0}\left[\cos \frac{2\pi  
x}{a}+1\right]$. Eq.~(\ref{ham}) is valid when the wells of this potential
are sufficiently deep: $U_{0}\gg \hbar ^{2}/ma^{2}$, where $m$ is the mass
of one atom. In this intense-laser regime, one can approximate the orbitals
at the bottom of each well by those of a harmonic oscillator, and the usual
method \cite{Ashcroft-Mermin-76} yields the tight-binding parameter as
\cite{Buchler-Blatter-Zwerger-02} $t=U_{0}\exp \left[-\frac{\pi}{2}
\sqrt{U_{0}\left(\hbar ^{2}/ma^{2}\right)^{-1}} \right],$ which is much
smaller than the first excitation energy of the orbital. The effect of the
overall trapping potential on $t$ can be neglected   provided that the
potential energy difference between the minima of neighbouring wells,   $\Delta E_j$, is
much less than the barrier height, $U_0$.  Now, $\Delta E_j \approx \kappa
a^2 j$ for $j \gg 1$;   hence our model is accurate for states whose wave
functions do not extend   beyond site $j_c \equiv U_0 / \kappa a^2$.  Hence
the state's energy $\ve$ must satisfy $\ve \ll \kappa a^2 j_c^2$, i.e.\
$\ve \ll \ve_{\rm max} \equiv U_0^2 / \kappa a^2$.  We would therefore like
to be sure that $\ve_{\rm max} \gg   t$, which is true provided that $U_0
\gg \kappa a^2$.

We are interested in the density of states (DOS),
$\rho(\ve) \equiv \sum_\nu \delta \left(\ve-\ve_\nu\right)$,
which is a smooth curve only in the limit in which the spectrum $\left\{  
\ve _{\nu }\right\} $
is a continuum. Let us discuss some limiting behaviours of $\rho(\ve)$.
The solution is well known for $\kappa =0$,
when the lattice forms a large {}`box' of length
$L$. In this case one obtains Bloch waves with the band dispersion  
relation
$
\ve _{k}=-2t\cos ka, \label{banddisp}
$
where $k$ is the wave number. When $L\gg a$,
$ka$ runs continuously from $-\pi $ to $\pi $.
In this limit, $\rho(\ve)$ is non-zero only for $\vert \ve \vert
\leqslant 2t$ (this region we call the `band'),
and has square-root singularities at $\ve=\pm
2t$ (the `top' and `bottom' of the band).
For $d=2$ the DOS extends from $\ve =-4t$
to $4t$ and is finite at these points, but exhibits a `van Hove'
logarithmic singularity at $\ve =0$ \cite{Ashcroft-Mermin-76}.

In our model, the lattice itself is taken to be infinite; the finite  
extent of the
wave functions is determined instead by the strength, $\kappa$, of the  
harmonic trap.
The corresponding length scale is  
\cite{Damle-Senthil-Majumdar-Sachdev-96} $l \equiv
\sqrt{t/\kappa}$.
This suggests that, in this model, `continuum limit' should be taken to  
mean
$l \gg a$
or, equivalently, $\kappa a^2 \ll t$. We shall work in this limit  
henceforth (unless otherwise stated).

Let us consider the low- and high-energy states of (\ref{ham}).
For energies near the bottom of the  
band, the
dispersion relation has the free-particle form $\ve_k \approx -2t
+ \hbar^2 k^2 / 2m^*$, where $m^* = \hbar^2 / 2 t a^2$. Thus we would expect the
low-lying eigenstates of (\ref{ham}) to resemble those of the usual  
continuum
harmonic oscillator.  Indeed, one can show explicitly that
$\left|\phi _{0}\right\rangle =N_{0}\sum
_{j}e^{-a j^{2}/2\sqrt{2}l}\left|j\right\rangle$
is the ground state to second order in $a/l$. Its energy is $\ve  
_{0}=-2t+\hbar \omega
^{*}/2+o\left([a/l]^2\right),$ where $\omega ^{*}\equiv \sqrt{\kappa  
/m^{*}}$. To first
order in $a/l$, this equals the usual result for the harmonic  
oscillator, measured from $\ve = -2t$.   The low-lying states are  
obtained similarly.
Since the DOS of the
1D harmonic oscillator is constant and equal to $1/\hbar \omega ^{*}$,
we conclude that the low-energy limit of the DOS of (\ref{ham}) is  
given by
$\rho \left(\ve \right)\rightarrow 1/\hbar \omega ^{*}$ as $\ve \to -2t$
for $d=1$.
For $d=2$, the DOS of the harmonic oscillator is $\rho \left(\ve  
\right)\propto \ve$,
so we expect a vanishing density of states as $\ve \to -4t$ in this  
case.

For high energies, the physical nature of the states is quite  
different. When $\ve \gg t$, the first term in (\ref{ham})
can be neglected, and thus the eigenstates of the Hamiltonian
are the position eigenstates  
$\left|j\right\rangle$, with energies $\ve _{j}=\kappa  
a^{2}j^{2}/2.$
The high-energy 1D DOS is then easily shown to be
$\rho(\ve \gg t) = \sqrt{2/\kappa a^2}\,\ve^{-1/2}$.
In 2D, the analogous calculation gives $\rho(\ve) \to 2\pi/\kappa a^2$ as $\ve/t \to \infty$.

\emph{WKB analysis.}
To extend our treatment of (\ref{ham}) to all energies, we employ the
Wentzel-Kramers-Brillouin (WKB) approximation \cite{Bender-Orszag-99}; it has already been
applied to the momentum-space
version of the problem in ref.~\cite{Polkovnikov-et-al-02}. By contrast, our WKB-type analysis will be performed in
real space \cite{ODell-01}.
It differs from traditional WKB
in that $\hbar$ does not appear
anywhere in (\ref{ham}); instead we choose $a$ as our small parameter.
As in the ordinary WKB method, we write the wave function in the form
\(
\psi(x) \sim \exp \left( i \int\limits^x k(x') dx' \right)
\). The energy is determined by the usual quantisation condition, $\oint k(x) \,dx = 2\pi \left( n + \gamma \right)$, where $n$ is an integer and $\gamma$
a constant. Differentiating this, we obtain the DOS: \( 
\rho(\ve) \equiv dn/d\ve = \frac{1}{2\pi} \frac{\partial }{\partial\ve}  \oint  
k(x) \,dx \) (note that $\gamma$ is not required). 

On the other hand, the orbit equation is modified by replacing the  
free-particle dispersion by the tight-binding
form \cite{Lifshitz-Azbel-Kaganov-73}:
$-2t \cos \left( ka \right) + \frac{1}{2} \kappa x^2 = \ve$. The local wavenumber becomes
$k(x) = \pm  
\frac{1}{a} \arccos \left(
\frac{\kappa x^2 - 2\ve}{4t} \right)$.
This makes a  
significant difference, because it introduces two new
turning points. The classical turning points (where  
$k=0$) still exist, and are found at $x=\pm x_c$,
where \be x_c = \sqrt{\frac{2\ve+4t}{\kappa}} = 2l \sqrt{  
\frac{\ve}{2t} +
1} \,\,;
\label{xc} \ee now, however, two new points appear for energies $\ve >
2t$.
They are associated with Bragg reflection, and they occur at $x = \pm  x_b$,
where \be x_b = \sqrt{\frac{2\ve-4t}{\kappa}} = 2l \sqrt{  
\frac{\ve}{2t} -
1}\,\,.
\label{xb} \ee
We shall refer to them as `Bragg turning points', and the region  
between them as `Bragg-forbidden'.
  
The quantisation condition now becomes more subtle.
To see why, note that in the Bragg-forbidden region $k(x)a = \pi + i \,{\rm arccosh} \left( \frac{2\ve - \kappa x^2}{4t}  
\right)$:\ the wave function
{\em continues to oscillate\/} for $\vert x \vert < x_b$, despite the 
damping of its amplitude.
The quantisation condition must
therefore include these oscillations; it becomes:
\be 2 \int\limits_{x_b}^{x_c}  
k(x)\,dx + \frac{2\pi x_b}{a} = \pi
\left( n + \gamma \right). \label{goodquant} \ee

Now we can obtain the DOS by differentiating (\ref{goodquant}),  
remembering that
$x_b$ and $x_c$ depend on $\ve$ according to (\ref{xc},\ref{xb}); we  
obtain a lengthy analytic expression for
$\rho(\ve)$.  It is plotted in  
Fig.~\ref{cap:Fig.1}(a), along with the DOS
obtained from numerical diagonalisation of (\ref{ham}),  
with appropriate provision for finite-size scaling (see below):\  
clearly the agreement is excellent.
This was foreseeable, since aside from a vanishing number of low-lying  
states, all the wave functions have slowly
varying $k(x)$ (or, in the Bragg-forbidden region, not varying at all);  
thus the WKB method is expected to be accurate
at all energies.  Note that $\rho(\ve)$ is {\it  
qualitatively} different from the DOS of the $\kappa=0$
tight-binding model, while it agrees {\it quantitatively} with
$\rho \to 1/\hbar \omega^*$ at $\ve=-2t$ and with  
$\rho \to \sqrt{2/\kappa a^2}\,\ve^{-1/2}$ for $\ve/t
\gg 1$, the limiting expressions obtained above.

Calculating the density of states for $d>1$ is  
simple, because the Hamiltonian is separable.
Consider, for example, the case of $d=2$.
%Let $H_{\rm 2D}$ be the Hamiltonian of the 2D system:\ it can be
%written $H_{\rm 2D} = H_x + H_y$, where $H_x$ and $H_y$ are two copies
%of the 1D Hamiltonian (\ref{ham}).  In such a situation, i
It is not hard to show that
$
\rho_{\rm 2D}(\ve) = \int_{-\infty}^\ve \rho(\ve-\chi)  
\rho(\chi) d\chi$,
where $\rho_{\rm 2D}$ is the DOS of the 2D system, and $\rho$ is the  
DOS of the 1D Hamiltonian (\ref{ham}).
The result
of applying this formula is presented in Fig.~\ref{cap:Fig.1}(b), and  
compared with numerical results; again, the
agreement is excellent.  Note that the steep feature at $\ve=0$ and the  
kink at $\ve=4t$ both result from the
logarithmic singularity in the 1D DOS.  As $\ve \to -4t$, we recover  
$\rho_{\rm 2D} \to 0$, as expected from the analytic
arguments above.  Furthermore, the large-$\ve$ asymptote is exactly the  
constant predicted by that analysis.
Higher-dimensional
densities of states may be generated in a similar way.

\emph{Finite-size scaling:\ numerics.} In the foregoing we obtained the
DOS in the continuum limit $l/a\rightarrow \infty $. However in
experiments $l/a$ is finite. We should therefore check that our
results resemble the behaviour of the system when it has a
steeper trap. Thus we have studied the single-particle spectrum and wave
functions by numerical diagonalisation of (\ref{ham}) \footnote{We
used the diagonalisation function of \cite{Octave}.}. By considering
increasingly shallow traps, the numerical calculations also allow us to
confirm the validity of WKB for the continuum
limit. We have applied the same method in $d=2$,
starting with the 2D version of (\ref{ham}). The results are analogous so
we shall, as above, describe in detail the 1D case only.

\begin{figure}
\includegraphics[keepaspectratio]{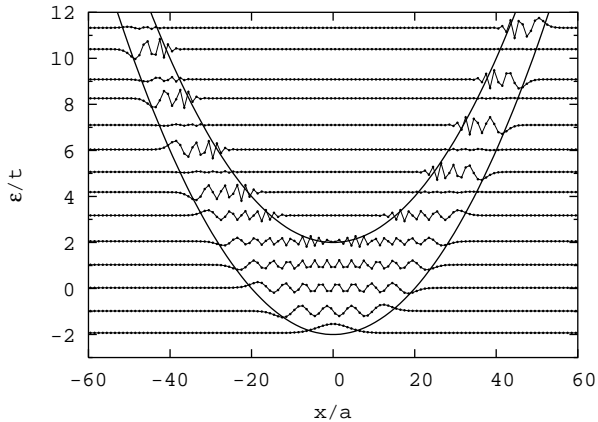}
\caption{\label{cap:Fig.3}Numerically determined wave functions for a  
trap
with $\kappa a^2=t/100$. Each wave function
has been offset along the $y$-axis by its energy, in units
of the hopping integral. The numerical calculation uses a lattice
with $N=130$ sites. For the energies considered, there are no noticeable
numerical artefacts. The lower and upper solid curves correspond to
the classical and Bragg turning points, eqs.~(\ref{xc},\ref{xb}).}
\end{figure}

\begin{figure}
\includegraphics{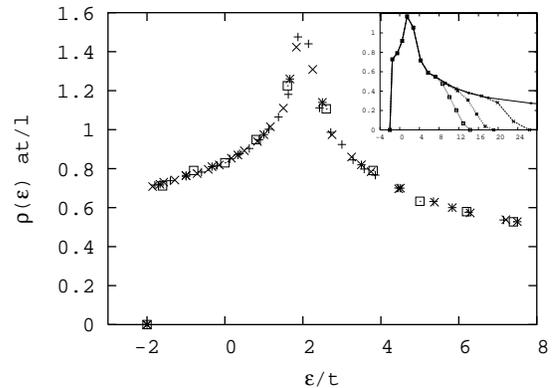}
\caption{\label{cap:Fig.4}Finite-size scaling of the binned DOS for 1D
lattices with finite strength of the trapping potential: $\kappa  
a^{2}/t=5 \cdot
10^{-6}\, (+),\, 10^{-5}\, (\times ),\, 10^{-4}\, (*)\textrm{ and  
}10^{-3}\, (\square )$.
In all cases the numerical cutoff energy, $\ve _{N}$, lies outside
the energy range showed in the plot. This is illustrated in the inset:\  
it
shows the DOS for fixed $\kappa  
a^2=t/1000$,
but calculated using lattices with different numbers of sites: $N=300\,  
(\square ),\, 350\, (*),\, 420\, (\times )\textrm{ and }2000\, (+)$. }
\end{figure}

\begin{figure}
\includegraphics{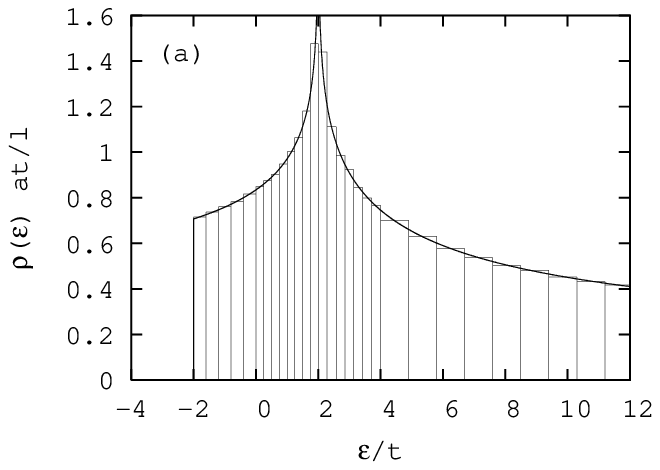}
\includegraphics{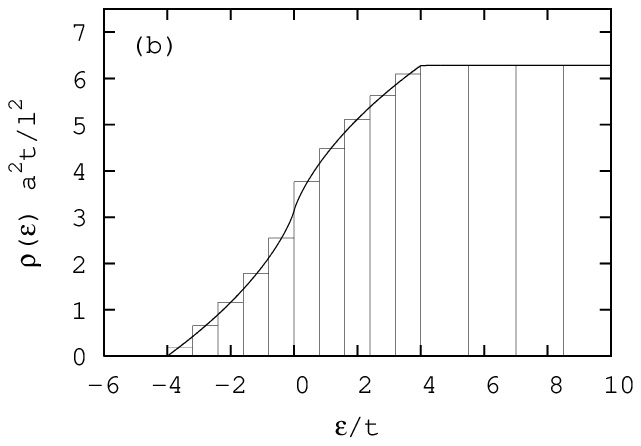}
\caption{\label{cap:Fig.1}DOS of a single atom in an optical lattice
for (a) $d=1$ and (b) $d=2$. Curves: continuum-limit results
obtained within the WKB approximation. Histograms: numerical
diagonalisation for finite-size systems with trap strengths $\kappa
=5\times 10^{-6}\, t/a^{2}$ (a) and $\kappa =1.5\times 10^{-2}\,
t/a^{2}$ (b), using lattices with $N=5000$ and $10000$ sites
respectively.}
\end{figure}

In order to have a finite-dimensional Hamiltonian matrix, we must describe
the lattice by a finite number of sites, $N$.  This leads to numerical
artefacts for $\ve >\ve _{N}$, defined by $N\equiv 2x_{c}(\ve _{N})/a$, as
the particle visits the artificial limits of the lattice. On the other
hand we expect this to be   essentially equivalent to the original problem
for $\ve \ll \ve _{N}$. Calculations for increasingly large
$N$ confirm this (see the inset of Fig.~\ref{cap:Fig.4}).

Fig.~\ref{cap:Fig.3} shows some numerically determined wave functions, for fixed trap strength. In spite of the relatively small system
size ($l/a=10$), the states seem well described
by the continuum-limit solution. At low energies, $\ve
\lesssim 0$, the wave functions resemble those of a continuum 1D harmonic
oscillator. At higher energies, the period of these oscillations starts to
become similar to the lattice constant, and commensuration effects emerge.
At $\ve =2t$, as that period reaches its minimum, the Bragg reflection
points appear. Immediately above the top of the band, $\ve \gtrsim 2t$, the
modulation of the wave number is evident, reaching its minimum and maximum
near the classical and Bragg turning points, respectively. The rapid  
oscillations continue in the Bragg-forbidden region, where the amplitude
decays exponentially. As the energy rises further, the distance between the
Bragg and classical turning points shrinks, so that fewer and fewer
oscillations take place between these two points. Eventually the particle is
forced to localise on single sites (not shown). Note that the Bragg and
classical turning points are very accurately described by
Eqs.~(\ref{xc},\ref{xb}), even for the relatively small value of $l/a$
considered here. Furthermore, the number of high-energy (localised)
solutions is reproduced correctly.  This is because the condition that the
wave function's phase be single-valued reduces, at $k=\pi$, to the condition
that the localised states be   separated by an integer number of lattice
spacings ($x=na$).

We now turn to the DOS. Obviously for finite $l/a$ the spectrum is
not a continuum so the DOS is not a smooth curve.
However, one can smooth it by the  
following
procedure:\ we divide an energy range into intervals, of width
$\Delta \ve $, and count the states within
each interval. We then re-scale the vertical axis of the
resulting histogram so that each column represents the number of states
per unit energy, on average, in the corresponding interval.
The result is a `binned' DOS that looks smooth only if the intervals are sufficiently wide. Fig.~\ref{cap:Fig.4} shows
the result for different values
of $\kappa a^{2}/t$. Note that we have further re-scaled the vertical
axis by an overall factor of $a/l$. This leads to the collapse of
all the data onto a single curve, suggesting that the continuum-limit
DOS describes the overall distribution of energy levels even for quite
steep traps, with only $l/a\sim 100\textrm{ sites}$. The inset  
illustrates
the numerical cutoff artefact mentioned above.
Finally, Fig.~\ref{cap:Fig.1} compares the DOS of two finite
(but fairly large) systems to the WKB predictions obtained above in
1D and 2D respectively. Evidently our WKB approach captures
the continuum limit rather well.

\emph{Conclusion.} We have calculated the DOS for a single atom in an
optical lattice; this should be regarded as the logical first step towards a
detailed theory of the experimentally realised many-particle systems.
Our results are based
on WKB theory, and refer to the limit $l \gg a$ or,
equivalently, when the trapping potential becomes flat: $\kappa
\rightarrow 0$. Numerical diagonalisation reveals this theory to be
extremely accurate in that case, and moreover shows that, for
finite-size systems, the binned DOS has the same overall features.
Our main result is that the DOS, in this limit, is radically
different from what is obtained for a homogeneous lattice (i.e.\ for
$\kappa=0$ rather than $\kappa \rightarrow 0$): the square-root
singularities in the 1D case are replaced by a
logarithmic one, and the logarithmic van Hove singularity in 2D
disappears altogether. Moreover, our theory provides a
detailed picture of how this comes about. The crucial feature is the  
new turning points, associated with Bragg
reflection, that appear at energies above the top of the conduction
band. The possibility of inducing Bragg reflection by using a  
time-dependent external potential was considered in  
\cite{Scott-Martin-et-al-03}. We have shown that Bragg reflection is,  
in fact, essential to understand the equilibrium single-particle  
spectrum of the optical lattice. 

\begin{acknowledgements}

We thank J.~M.~F.~Gunn, A.~F.~Ho, M.~W.~Long and A.~J.~Schofield for useful discussions. We acknowledge financial support from EPSRC (CH) and the Leverhulme Trust (JQ).

{\it Note added.---} It has been drawn to our attention that some of the features discussed here have recently been pointed out in an exact diagonalisation study \cite{Rigol-Muramatsu-03}.
\end{acknowledgements}

\bibliographystyle{apsrev}
\bibliography{bibliography}

\end{document}